\documentclass[12pt,preprint]{aastex}

\newcommand{\myemail}{cmorgan@usna.edu}
\newcommand{\kbar}{\langle\kappa\rangle}

\shorttitle{MICROLENSING IN SDSS 0924+0219}
\shortauthors{MORGAN ET AL.}

\begin{document}

\title{Microlensing of the Lensed Quasar SDSS 0924+0219\footnote{Based on 
  observations obtained with the Small and Moderate Aperture Research Telescope System
  (SMARTS) 1.3m, which is operated by the SMARTS Consortium,
   the Apache Point Observatory 3.5-meter telescope, which is 
  owned and operated by the Astrophysical Research Consortium, the WIYN Observatory
which is owned and operated by the University of Wisconsin, Indiana University, 
Yale University and the National Optical Astronomy Observatories (NOAO), the 
6.5m Magellan Baade telescope, which is a collaboration between the observatories 
of the Carnegie Institution of Washington (OCIW), University of Arizona, 
Harvard University, University of Michigan, and Massachusetts Institute 
of Technology, and observations made with the NASA/ESA Hubble Space Telescope
  for program HST-GO-9744 of the Space Telescope Science Institute,
  which is operated by the Association of Universities for Research
  in Astronomy, Inc., under NASA contract NAS 5-26555.
}}

\author{Christopher W. Morgan\altaffilmark{2}}
\affil{Department of Physics, United States Naval Academy, 572C Holloway Road,
Annapolis, MD 21402}
\email{\myemail}
 
\author{C.S. Kochanek and Nicholas D. Morgan}
\affil{Department of Astronomy, The Ohio State University, 140 West 18th Avenue, Columbus, OH 43201-1173}
\email {ckochanek@astronomy.ohio-state.edu, nmorgan@astronomy.ohio-state.edu}

\and

\author{Emilio E. Falco}
\affil{Harvard-Smithsonian Center for Astrophysics, 60 Garden Street, Cambridge, MA, 02138}
\email{efalco@cfa.harvard.edu}

\altaffiltext{2}{Department of Astronomy, The Ohio State University}

\def\avgm{\langle M \rangle}
\def\avgmhat{\langle M/M_\odot \rangle}

\begin{abstract}
We analyze V, I and H band HST images and two seasons of $R$-band monitoring data for the 
gravitationally lensed quasar SDSS0924+0219.  We clearly see that image D is a point-source
image of the quasar at the center of its host galaxy.  We can easily track the host 
galaxy of the quasar close to image D because microlensing has provided a natural 
coronograph that suppresses the flux of the quasar image by roughly an order of magnitude. 
We observe low amplitude, uncorrelated variability between the four quasar images
due to microlensing, but no correlated variations that could be used to measure a
time delay.  Monte Carlo models of the microlensing variability provide estimates
of the mean stellar mass in the lens galaxy ($ 0.02 {\rm M_\sun} \lesssim 
\avgm \lesssim 1.0 {\rm M_{\sun}}$),
the accretion disk size (the disk temperature is $5 \times10^4$~K at 
$3.0 \times 10^{14}\:{\rm cm} 
\lesssim r_s \lesssim 1.4 \times 10^{15} \: {\rm cm}$), and the black hole mass 
($2.0 \times 10^7 {\rm M_{\sun}} \lesssim M_{BH}\, \eta_{0.1}^{-1/2} \, (L/L_{E})^{1/2} 
\lesssim 3.3 \times 10^8 {\rm M_{\sun}}$), all at 68\% confidence. The black hole
mass estimate based on microlensing is consistent with an estimate 
of $M_{BH} =(7.3 \pm 2.4 )\times 10^7 {\rm M_{\sun}}$ from the \ion{Mg}{2} emission line width.  
If we extrapolate the best-fitting light curve models into the future, 
we expect the the flux of images A and B to 
remain relatively stable and images C and D to brighten.  In particular,
we estimate that image D has a roughly 12\% probability of brightening by a 
factor of two during the next year and a 45\% probability of brightening by
an order of magnitude over the next decade.
\end{abstract}


\section{Introduction}

In the standard Cold Dark Matter (CDM) galaxy model, 
early-type galaxies are composite objects. Their mass is dominated by an extended dark
matter halo that surrounds the luminous stars of the visible galaxy; any remaining baryons are
left as hot gas \citep{White1978}.  The halos grow by mergers with other halos,
with a small fraction of the accreted smaller halos surviving as satellite
halos (``sub-structure") orbiting in the larger halos,  
but the mass fraction in these satellites is uncertain
(\citealt{Moore1999}, \citealt{Klypin1999}, 
\citealt{Gao2004}, \citealt{Taylor2005} and  \citealt{Zentner2005} ).  
There is increasing evidence from time delay measurements
(e.g. \citealt{Kochaneketal.2005}) and stellar  dynamical observations
(e.g. locally, \citealt{Romanowsky2003}, and
in lenses \citealt{Treu2002}, \citealt{Treu2006}), that the density structure of some
early-type galaxies on scales of 1--2$R_e$ is heterogeneous, but this needs to be changed from a
qualitative assessment to something more quantitative.
Thus the detailed balance between stars, dark matter and 
substructure (luminous or dark) remains a matter of debate.

One approach to addressing these problems is to monitor 
variability in gravitational lenses. Variability in lenses arises from two sources.  
Correlated variability between the images due to fluctuations in the source flux 
allows the measurement of the time delays $\Delta t$ between the quasar images,
which constrain the combination  $\Delta t \propto (1-\kbar)/H_0$ of the surface
density near the lensed images $\kbar=\langle\Sigma\rangle/\Sigma_c$ and the
Hubble constant $H_0$ to lowest order \citep{Kochanek2002}.\footnote{The dimensionless
surface density $\kappa$ is the surface density $\Sigma$ divided by
the critical surface density $\Sigma_c\equiv c^2 D_{\rm OS}/4\pi G
D_{\rm OL}D_{\rm LS}$, where $D_{\rm OL}$, $D_{\rm OS}$ and $D_{\rm
LS}$ are the angular diameter distances between the Observer, Lens and
Source.}  By measuring the surface density near the lensed images, we can
strongly constrain the radial mass profile of the lens.  
Uncorrelated variability between the images is a signature 
of microlensing by stars in the lens galaxy.  Microlensing can constrain
the mass distribution because the 
statistics of the variations depend on the fraction of the local density
in stars \citep{Schechter&Wambsganss2002}.  
In addition to providing an estimate of the surface density in 
stars near the lensed images, microlensing can also be used to estimate the 
mean stellar mass in the lens and to determine the structure of quasar accretion 
disks.  Understanding microlensing is also required to improve estimates of the
substructure mass fraction in radio--quiet lenses, because most other
quasar source components are affected by both substructure and microlensing.
In order to use time variability in lenses to probe these astrophysical problems, 
we have undertaken a program to monitor roughly 25 lenses in several optical and near-IR
bands.  The first results of the program and a general description of our
procedures are presented in \citet{Kochaneketal.2005}. 

In this paper we study the four-image $z_s=1.52$ quasar lens SDSS 0924+0219 \citep{Inada2003}
using $V$, $I$ and $H$-band Hubble Space Telescope ({\it HST}) observations and the results from
two seasons of monitoring the system in the $R$-band.
The lens galaxy is a fairly isolated $z_l=0.393$ \citep{Ofek2005}
early-type galaxy.  The most remarkable feature of this lens is that it shows a spectacular
flux ratio anomaly between the A and D images.  These two images are merging at a fold
caustic with a flux difference of nearly 3 magnitudes when they should have approximately
equal fluxes by symmetry \citep{Keeton2005}.  
\citet{Keeton2006} recently measured the flux
ratios in the Ly$\alpha$ line and the adjacent continuum, finding that the anomaly
is weaker in the emission line but still present.  This indicates that the anomaly
is partly due to microlensing, since the expected size difference between
the broad line region and the optical continuum emission region should matter for
microlensing but be irrelevant for substructure.  Unfortunately, the 
continued existence of the anomaly in the emission line means either that the broad line 
region is not large enough to eliminate the effects of microlensing or that the
flux ratio anomaly in SDSS0924+0219 is due the combined effects of microlensing
and substructures.  

We present the HST data in \S\ref{sec:observations} as well as a series of mass models 
for the system consisting of the observed stellar distribution embedded in a standard
dark matter halo.  In \S\ref{sec:models} we present the light curves and model them
using the Monte Carlo methods of \citet{Kochanek2004}.  The analysis allows us to
estimate the mean stellar mass in the lens galaxy, the size of the quasar accretion 
disk and the mass of the black hole powering the quasar.  In
\S\ref{sec:expectations} we present our predictions for
the expected variability of this source over the next decade.
Finally, we summarize our findings in \S\ref{sec:conclusions}. All calculations in this paper
assume a flat $\Lambda$CDM cosmology with $h=0.7$, $\Omega_{M}=0.3$ and $\Omega_{\Lambda}=0.7$.

\section{HST Observations and Mass Models}
\label{sec:observations}

In this section we discuss the HST observations and our mass model fits to
the astrometric and photometric measurements.

\subsection{HST Observations}
\label{sec:hst}
\begin{figure}
\epsscale{1.0}
\plotone{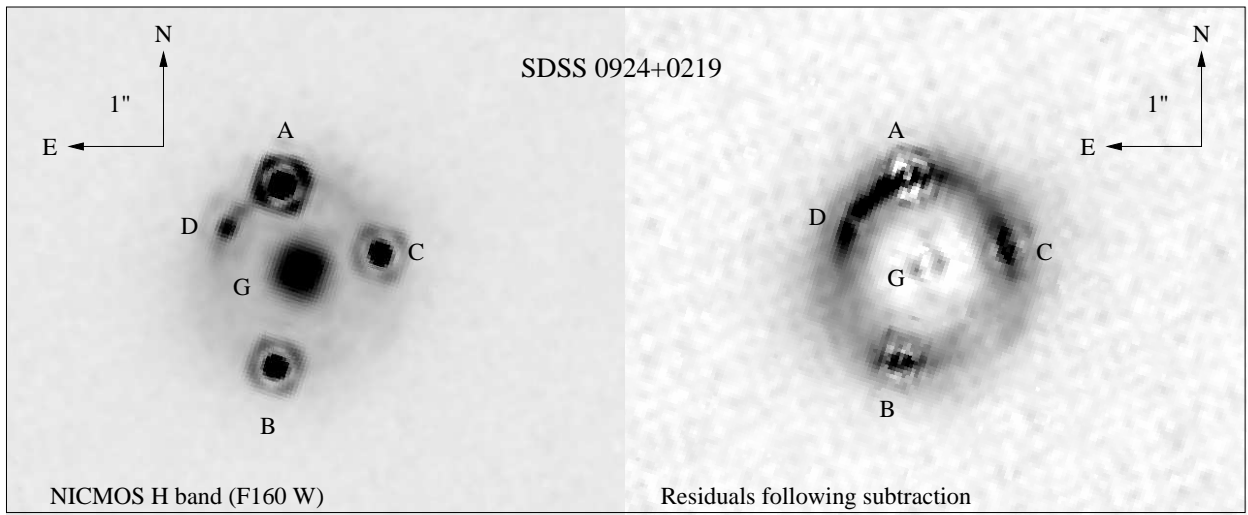}
\caption{$H$-band images of SDSS0924+0219.  The left panel shows the original image.  Note that
image D shows the Airy ring of a point source and is markedly fainter than image A. 
The right panel shows $H$-band residual near each quasar image after 
subtracting the four quasar images and 
the lens galaxy to make the Einstein ring image of the quasar host galaxy more visible. 
The noisy residuals near each image are largely due to small errors in the PSF models.
\label{fig:hst}}
\end{figure}

We observed the lens in the $V$- (F555W), $I$- (F814W) and $H$-bands (F160W) using
{\it HST}. The $\sim4380$ sec $V$-band and the $\sim4600$~sec  
$I$-band images were obtained as 
eight dithered sub-images with the Wide Field Channel (WFC) of the Advanced Camera for 
Surveys (ACS) on 2003 November 18. The 5120~sec $H$-band 
image was obtained as eight dithered sub-images
on 2003 November 23 using the Near-Infrared Camera and Multi-Object Spectrograph 
(NICMOS).  
The ACS data were reduced using the {\it pyraf multidrizzle} package, 
and the NICMOS data were reduced using {\it nicred} \citep[see][]{Lehar2000}.
We focus on the results from the new $H$-band image,
shown in Fig.~\ref{fig:hst}. The four quasar images, the central lens galaxy and an 
Einstein ring image of the quasar host galaxy can easily be seen.

We fit the $H$-band image using a photometric model consisting of four point sources 
and a de Vaucouleurs model for the lens galaxy. We modeled the lensed host galaxy
as an exponential disk, using a singular isothermal ellipsoid (SIE) 
mass model for the lens galaxy. We also tried a de Vaucouleurs lens galaxy 
mass model, but doing so caused no significant change 
in the quality of the lensed host galaxy photometric fit (see \citealt{Peng2006}
for a discussion of systematic issues in modeling lensed host galaxies).
The fits were done with {\it imfitfits}  \citep[see][]{Lehar2000}
using a range of bright PSF models, with the PSF
producing the best overall fit being adopted for the final results.  
We then use the $H$-band fit as our reference, and we hold the astrometry and
model structure fixed for the $V$- and $I$-band photometric fits.
In Table ~\ref{tab:hstastrom_phot}, we 
present astrometric and photometric measurements for the system.
Our fits to the lens astrometry and photometry are consistent 
with those of \citet{Keeton2006} and \citet{Eigenbrod2005}. 

The lens in SDSS 0924+0219 is an early--type galaxy with effective
radius $R_e = 0\farcs31 \pm 0\farcs02$, axis ratio $q=0.92\pm0.02$, 
major axis position angle $\theta_{e}=-27 \pm 8 \degr$(East of North) 
and colors $V-I=1.44\pm0.07$ and $V-H=3.60\pm0.05$. Following the technique of
\citet{Rusin2005}, we performed a Fundamental Plane (FP) 
analysis of the lens to determine if its observed properties
fall on the track of the expected Mass--to--Light (M/L) evolution with redshift 
for elliptical galaxies \citep{Rusin2003}. Assuming that a 
nominal early--type galaxy undergoes a starburst phase
at high redshift and then evolves passively thereafter, we expect  
elliptical galaxies to show a steady increase in M/L with decreasing redshift. 
If the SDSS0924+0219 lens galaxy is to lie on the present-day FP, then it requires a
$\Delta \log(M/L)$ evolution of $-0.42\pm0.03$ from $z=0.393$ to $z=0$. This
is steeper than the mean value for lens galaxies around that redshift,
$\Delta \log(M/L) = -0.22\pm0.04$, implying that the lens has a smaller 
than average M/L. Note, however, that \citet{Treu2006} find a faster evolution than
\citet{Rusin2003}, which would be consistent with the value we find 
for SDSS0924+0219.

One remarkable feature in this lens is how the stars (or a small satellite)
in the lens galaxy have provided a natural coronograph at the location of 
image D.  The quasar flux is suppressed by roughly an order of magnitude, making
it very easy to see into the central regions of the host galaxy.  We obtain
a good fit with a host having a scale length of $0\farcs11\pm0\farcs01$, 
axis ratio of $0.74\pm0.05$ and magnitude of $H=20.56\pm0.14$~mag.  
For comparison, we estimate an unmagnified magnitude of $H=20.40\pm0.20$~mag 
for the quasar. We also attempted a de Vaucouleurs model fit to the host
galaxy, but we found that doing this resulted in a negligible 
change in the overall quality of fit.  We are not able to 
discriminate between host galaxy models.    
We extracted the Einstein ring curve of the lens \citep{Kochaneketal.2001}
to use as one of the constraints on the mass models.

Images A, B and C have similar $I-H$ and $V-I$ colors, while image D is significantly redder 
in $I-H$ but of similar color in $V-I$.  This is similar to the expected pattern for dust 
extinction, since an $0.5$~mag difference in $I-H$ should correspond to a  $0.15$~mag 
difference in $V-I$ for an $R_V=3.1$ extinction curve shifted to the redshift of the lens.
On the other hand, the Einstein ring image of the host galaxy does not show color trends
near the D image, which strongly argues against dust as the origin of the
color differences.  Moreover the significant differences between the A/D flux
ratio in Ly$\alpha$ as compared to the continuum \citep{Keeton2006}, means that microlensing
must be a significant contributor to the anomalous flux ratio.

\subsection{Macro Models and Substructure}
\label{sec:macromodels}

We modeled the lens galaxy as the sum of a de Vaucouleurs model with scale length $R_{e}=0\farcs31$
embedded in an NFW \citep{Navarro1996} model with a break radius $r_{c}=10\farcs0$.  The de Vaucouleurs
and NFW models were ellipsoids constrained by the axis ratio and orientation of the
lens galaxy in the $H$-band images, and we included an external shear to model any additional
perturbations from the lens environment or along the line of sight.  We constrained the mass model 
with the astrometry of the quasar images and the Einstein ring curve derived from the $H$-band images 
\citep{Kochaneketal.2001} using the GRAVLENS \citep{GRAVLENS} software package.
We required the NFW and de Vaucouleurs components to be perfectly concentric, but 
we permitted the combined model to move within $0\farcs01$ of the measured
galaxy center in order to optimize the fit.    
As described in detail by \citet{Kochanek2005}, 
it is not possible to determine the radial mass profile of the lens using these constraints,
although it can be done using other constraints such as a time delay or stellar velocity dispersion 
measurement.  Given this degeneracy, we generated a 
sequence of models parameterized by $0\leq f_{M/L}\leq1$, the 
fractional mass of the de Vaucouleurs component compared to a constant M/L model ($f_{M/L}=1$).
As expected, there is no significant difference in how well models with $0< f_{M/L}\leq 1$
fit the constraints with the exception of pure dark matter models ($f_{M/L}\lesssim 0.1$)
that predict a detectable, fifth or odd quasar image near the center of the Einstein ring. The convergence, shear and $\kappa_{*}/\kappa$ for the range of $f_{M/L}$ at 
each image location is presented in Table~\ref{tab:macromodels}.

\begin{figure}
\epsscale{1.0}
\plottwo{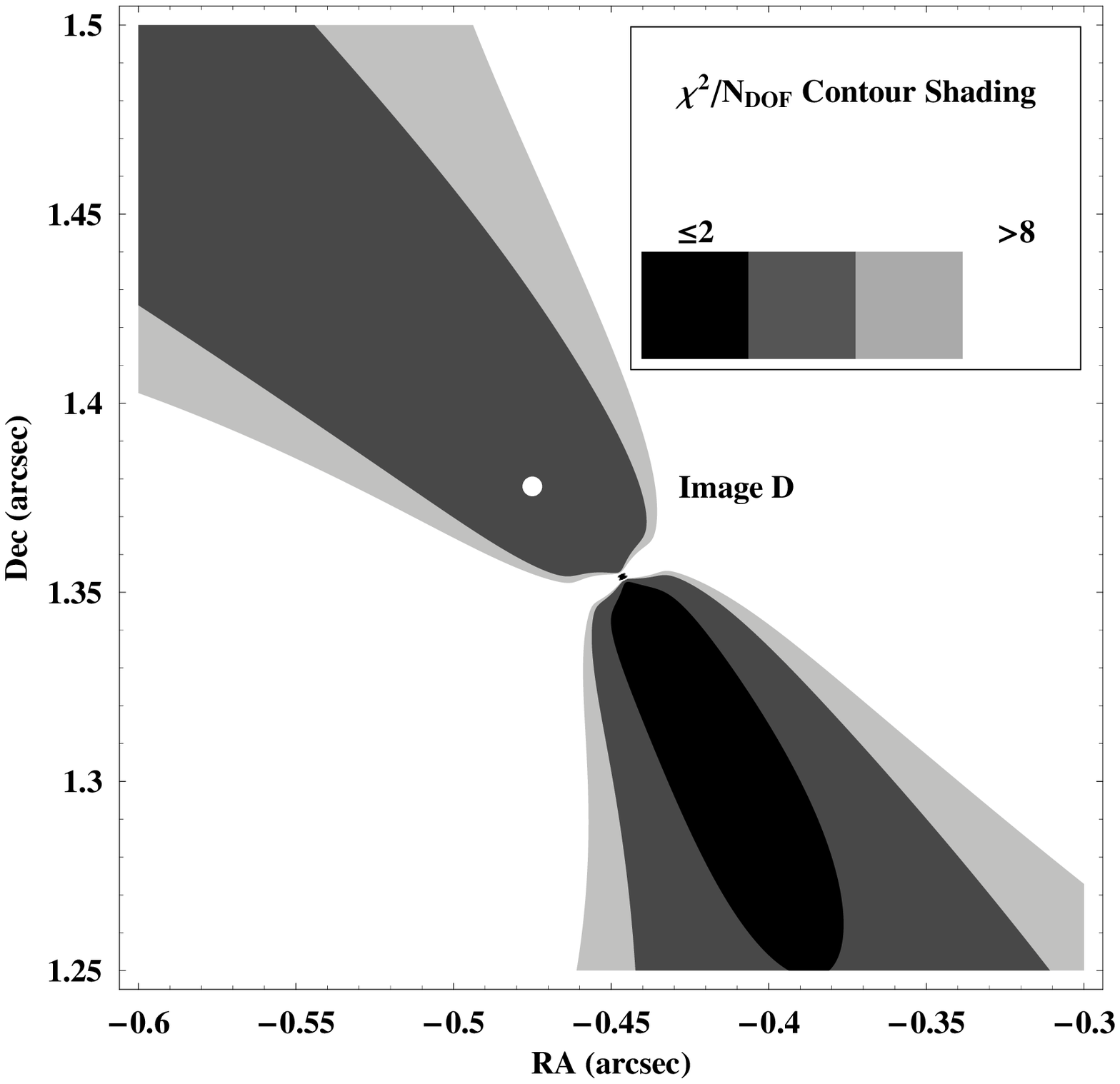}{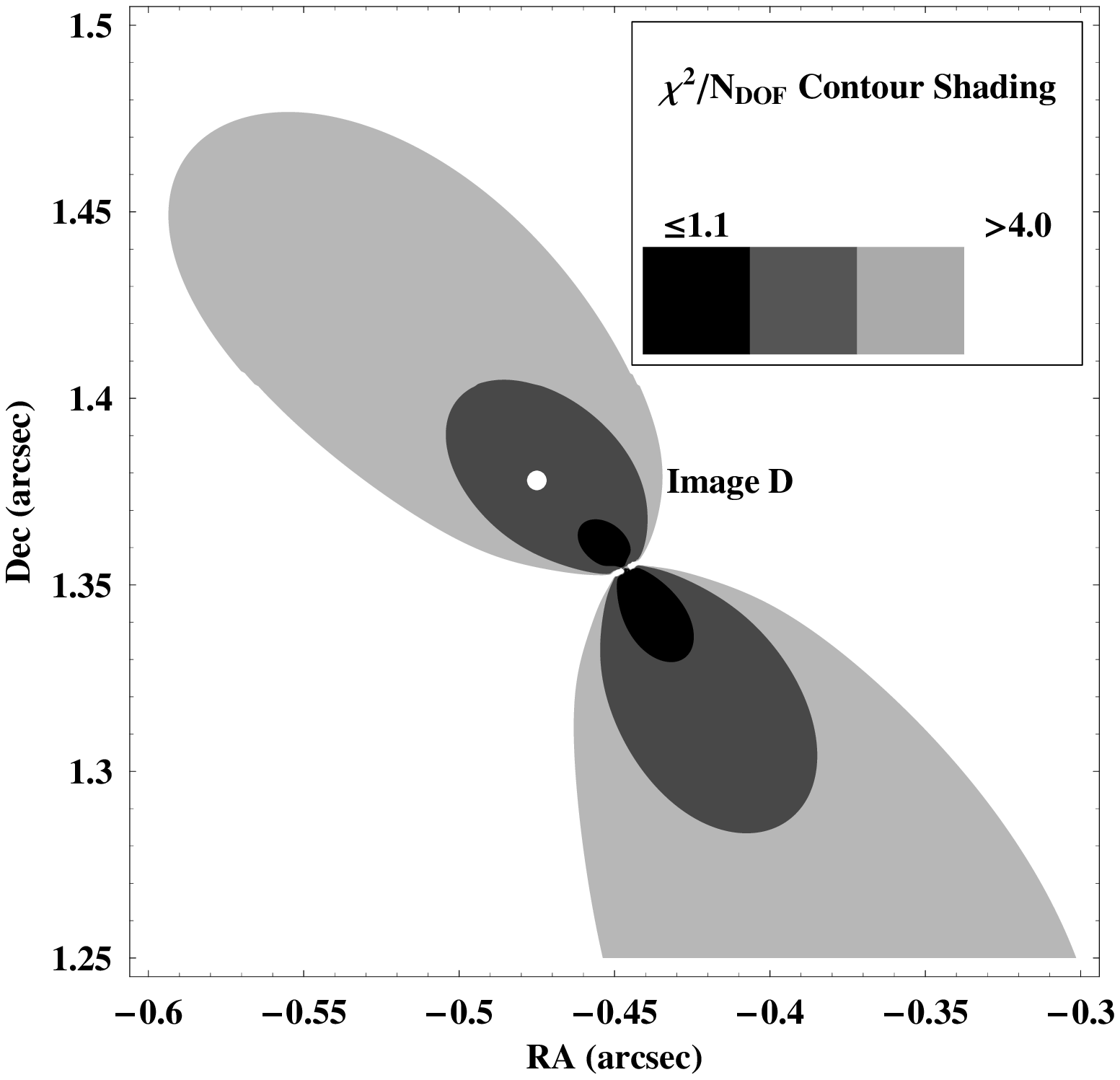}
\caption{Substructure models for the SDSS0924+0219 flux ratio anomaly.  The left (right) panels
show the regions where a pseudo-Jaffe model with an Einstein radius of 0$\farcs01$ ($0\farcs003$)
can remove the A/D flux ratio anomaly without significantly worsening the constraints on
the quasar image positions or the Einstein ring. The main lens is modeled as a singular
isothermal ellipsoid (SIE) and the pseudo-Jaffe models are tidally truncated at
$0\farcs1$ ($0\farcs06$). The position of image D is indicated with a white point.
$\chi^2$ contours are indicated in the figure legends.
\label{fig:substructure}}
\end{figure}

Although much of the anomalous A/D flux ratio must be due to microlensing based on the smaller
anomaly observed in the emission lines, we explored the extent to which the anomaly could
be created by small satellites of the lens galaxy rather than by stars.  For these tests we
modeled the main lens as a singular isothermal ellipsoid (SIE) and then added a low-mass
pseudo-Jaffe model as a perturber.  We assigned the perturber Einstein radii of either
$0\farcs01$ or $0\farcs003$ and tidal truncation radii of $0\farcs1$ and $0\farcs06$, respectively. 
Fig.~\ref{fig:substructure} shows the
goodness of fit, $\chi^2/N_{DOF}$, as a function of the perturber's position, where we fit
both the astrometric constraints from the quasar image positions and the flux
ratios from the $H$-band {\it HST} data.   While the more massive satellite
with an Einstein radius of 0\farcs01 has difficulty adjusting the flux ratio without
violating the astrometric constraints, the lower mass satellite can do so if properly
positioned.  Such a satellite would have a mass of $\sim 10^{-5}$ that of the primary
lens, roughly corresponding to the mass scale of globular clusters.  While the emission
line flux ratios largely rule this out as a complete explanation for SDSS 0924+0219, substructure
could explain the continued existence of an anomaly in the emission line flux ratios. We include
this calculation as an illustration that substructure can lead to 
anomalies as extreme as are observed here.  

\section{Microlensing}
\label{sec:models}

\subsection{Monitoring Data and Microlensing }
\label{sec:monitoring}

We have obtained somewhat more than two seasons of $R$-band 
monitoring data for SDSS 0924+0219.  Our analysis
procedures are described in detail in \citet{Kochaneketal.2005}, so we provide
only a brief summary here.  We measure the flux of each quasar image relative to
a sample of reference stars in each frame.  We keep the relative positions of 
the components fixed, using the {\it HST} astrometry for the lensed components, and
derive the PSF model and quasar flux for each epoch by simultaneously fitting
the lens and the reference stars.  The PSF is modeled by 3 nested, elliptical
Gaussian components. The galaxy is included in the model at a constant flux
which we determine by fitting all the data as a function of the galaxy flux
and then adopting the galaxy flux that produces the best fit to the complete
data set. We confirm that the lens galaxy flux is approximately constant 
at each epoch by examination of the residual galaxy flux after subtraction 
of the best--fit models, and we find no evidence for variability 
during our three seasons of monitoring.

\begin{figure}
\epsscale{1.0}
\plotone{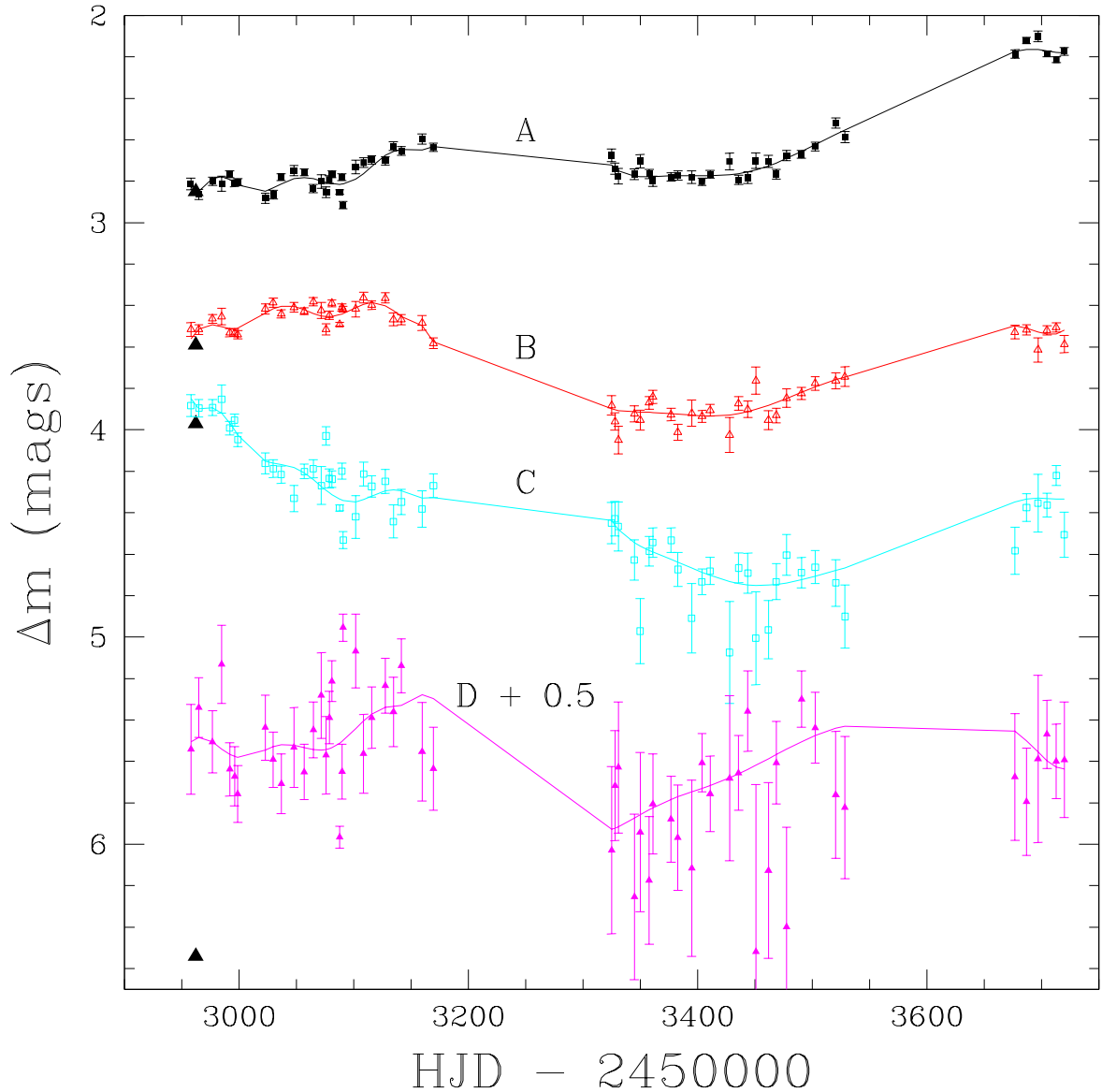}
\caption{SDSS0924+0219 $R$-band light curves for images A--D. The data points 
for image D are offset to improve their visibility. The curves are derived from a
joint, high-order polynomial fit for the source light curve combined with
lower order polynomials for the 
microlensing variability of each image
(see \citealt{Kochanek2005} for details). 
Symbols: Image A--solid squares, Image B--open triangles,
Image C--open squares and Image D--solid triangles.  The large solid triangles at 2962 days
are the HST $V$-band photometry referenced to Image A. 
\label{fig:lightcurves}}
\end{figure}

Most of our observations were obtained at the queue-scheduled SMARTS 1.3m using 
the ANDICAM optical/infrared camera 
\citep{Depoy2003}.\footnote{http://www.astronomy.ohio-state.edu/ANDICAM/}  
Additional observations were obtained at the Wisconsin-Yale-Indiana (WIYN) observatory
using the WIYN Tip--Tilt Module (WTTM) \footnote{http://www.wiyn.org/wttm/WTTM\_manual.html},
the 2.4m telescope at the MDM Observatory using the MDM Eight-K
\footnote{http://www.astro.columbia.edu/~arlin/MDM8K/}, Echelle and RETROCAM
\footnote{http://www.astronomy.ohio-state.edu/MDM/RETROCAM} \citep{Morgan2005}
imagers, the 3.5m APO
telescope using Spicam and the 6.5m Magellan Baade telescope using IMACS \citep{Bigelow1999}.
Images taken under seeing conditions worse than $1\farcs5$ were discarded.
We also added the photometry from \citet{Inada2003} to extend our baseline to nearly
four years for the microlensing calculations.  The $R$-band light curves are displayed in 
Fig.~\ref{fig:lightcurves}, and the data are presented in Table ~\ref{tab:lightcurves}. 

In Fig.~\ref{fig:lightcurves}, we also show the {\it HST} $V$-band photometry scaled to the
best--fit $R$-band monitoring magnitude of Image A on the observation date.  In the {\it HST}
data, image D is $\sim1$~mag fainter relative to image A than our estimate
on nearly the same date.  After considerable experimentation, we concluded
that our flux for image D may be contaminated by image A, although we
found no correlation between the A/D flux ratio and the seeing. Nonetheless,
we include estimates of seeing at each observation 
epoch in Table~\ref{tab:lightcurves}.  In the
calculations that follow we will use both our image D light curve as
observed and an image D light curve shifted 1~mag fainter to agree with the {\it HST} flux ratio.
The shift had little effect on our results in \S~\ref{sec:micromodels}--\ref{sec:expectations}.

In our mass models, the longest expected delay for the system is $\sim 10$~days for
$H_0=72$~km~s$^{-1}$~Mpc$^{-1}$.  We see little evidence for correlated variability
between the images on these time scales, so we cannot measure the time delays.  For
the present study, it seems safe to simply ignore the time delays. 

\subsection{Microlensing Models}
\label{sec:micromodels} 

For each of our macro models, parameterized by $f_{M/L}$, we generated 8 random realizations of 
the expected microlensing magnification patterns for each image.  We used patterns with an 
outer dimension of $20 R_e$ where 
$R_e = 5.7 \times 10^{16} \,  \langle M/M_{\sun} \rangle^{1/2}$~cm
is the Einstein radius for the average mass star.  The map dimensions were $8192^2$,
so we can model source sizes down to $3 \times 10^{-3}R_e$.  
The stars used to create the patterns were drawn from a Salpeter IMF with
a dynamic range in mass of a factor of one hundred.
We modeled the accretion disk of the 
quasar as a standard, face-on thin disk model \citep{Shakura1973} with a scale length
of $r_s = \hat{r}_s \langle M/M_\odot\rangle^{1/2}$ where the microlensing
behavior is determined by the source size scaled by the mean mass
of the microlenses, $\hat{r}_s$.  We have chosen to use a thin disk model because
it provides a context for interpreting the results, but \citet{Mortonson2005} have
shown that $r_s$ can simply be interpreted as the typical half-light radius for
any choice of emission profile. We do, however, neglect the central hole in the disk
emissivity to avoid the introduction of an additional parameter.   
We fit the light curves using the Monte Carlo method of \citet{Kochanek2004}.  
In this method, large numbers of trial light curves are randomly generated and fitted to the 
observed light curves. Bayesian statistical methods are used to combine the resulting
distributions of $\chi^2$ values for the fits to the light curves to obtain probability
distributions for the model parameters.   

\begin{figure}
\epsscale{1.0}
\plotone{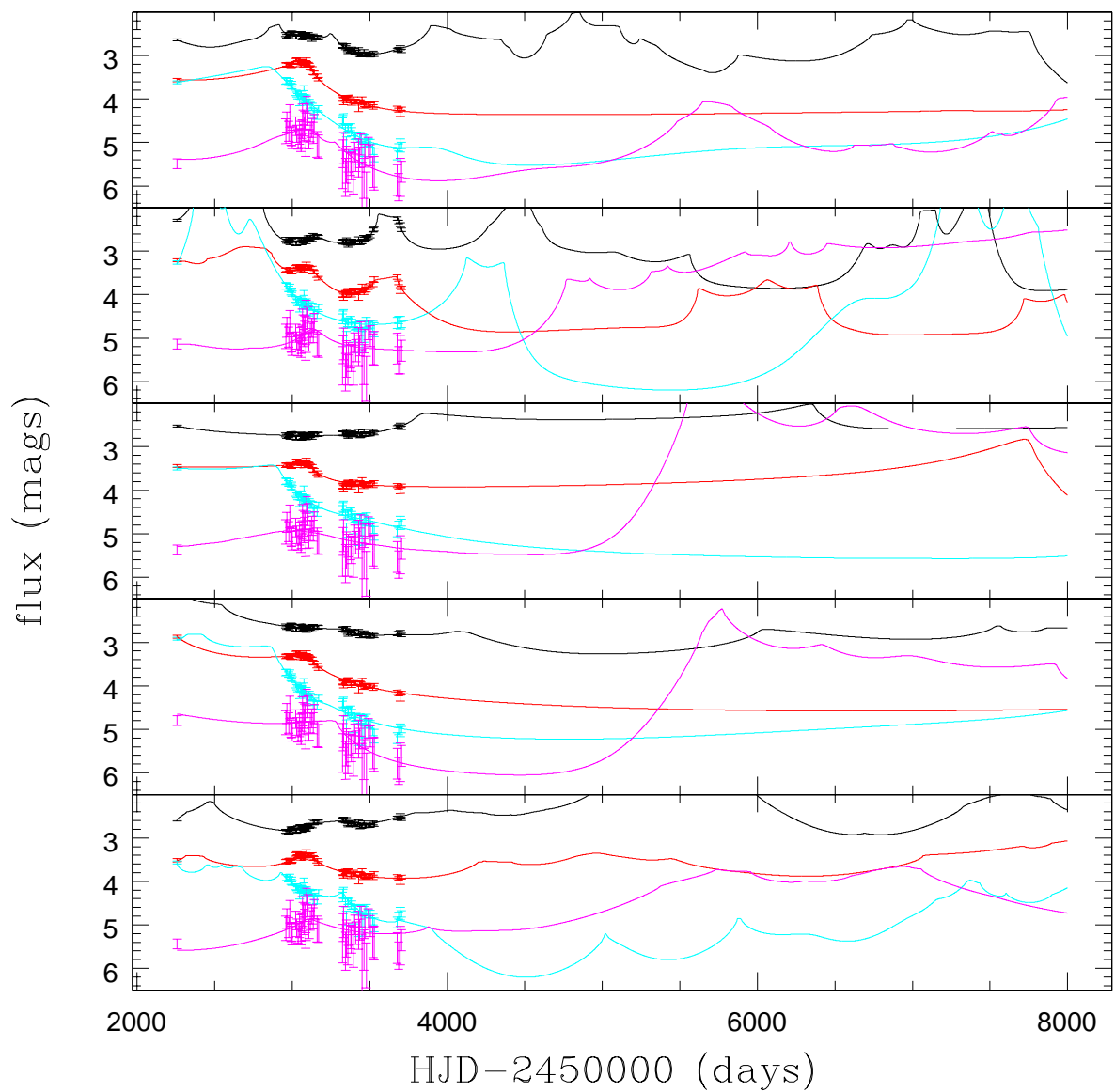}
\caption{The five best fitting $R$-band light curves extrapolated 
for an additional 10 years across their 
magnification patterns. 
In order to show the ``goodness of fit'' of the theoretical microlensing lightcurves, 
we plot as points the observed image flux minus our model for the intrinsic 
variability of the source.  In most of these
light curves we see a brightening of images D and C.
\label{fig:modellightcurves}}
\end{figure}

We are interested in models where microlensing is responsible for any deviations of the
image flux ratios from the lens model, so we assumed that the flux ratios of the macro model
were correct up to a systematic uncertainty of 0.05 mag for images A--C and 0.1 mag for 
image D.  We also allowed for an $0.02$~mag systematic uncertainty in the photometry of 
images A--C and an $0.1$~mag systematic uncertainty in the photometry of image D.  These 
errors were added because the point--to--point scatter in the light curves is somewhat broader 
than the formal uncertainties in the photometry.  With these assumptions, we have no
difficulty finding light curves that fit the data well, with $\chi^2/N_{DOF}\approx1.0$.  
We generated $10^5$ trial light curves for each source size, magnification pattern 
and mass model.  Several example light curves which provide good fits to the data are shown in 
Fig.~\ref{fig:modellightcurves}.  For each reasonable fit to the light curves, defined 
by $\chi^2/N_{DOF} \leq 2.3$, we also generated an extrapolated light curve extending for an 
additional ten years beyond the last data point of the third season (2005 December 14; HJD 2453719).
We also repeated all the calculations shifting the image D light curve 1~mag fainter to
match the HST observations, finding few changes in the results.

In order to convert the results from Einstein radius units, where all physical scales depend
on the mean mass of the microlenses $\langle M/M_\odot\rangle$, we must assume either
a probability distribution for the actual velocities or a prior for the mean stellar
mass.  Our velocity model includes the 176~km~s$^{-1}$ projected velocity of the 
CMB dipole onto the plane of the lens, a probability distribution for the one--dimensional peculiar
velocity dispersion of galaxies at $z_l$ with rms value of 164~km~s$^{-1}$ and a one--dimensional 
stellar velocity dispersion in the lens galaxy of 219~km~s$^{-1}$ based on the Einstein radius of
the lens and an isothermal lens model. As discussed in detail by \citet{Kochanek2004}, the lens galaxy 
peculiar velocity dispersion and stellar velocity dispersion estimates are dependent upon the selected cosmology.  We have chosen the standard flat concordance cosmology ($\Omega_{M}=0.3$
and $\Omega_{\Lambda}=0.7$) 
for our estimates. We also consider the consequences of using
a limited range for the mean stellar mass of $0.1 \leq 
\langle M/M_\odot \rangle < 1.0 $. 

\begin{figure}
\epsscale{1.0}
\plotone{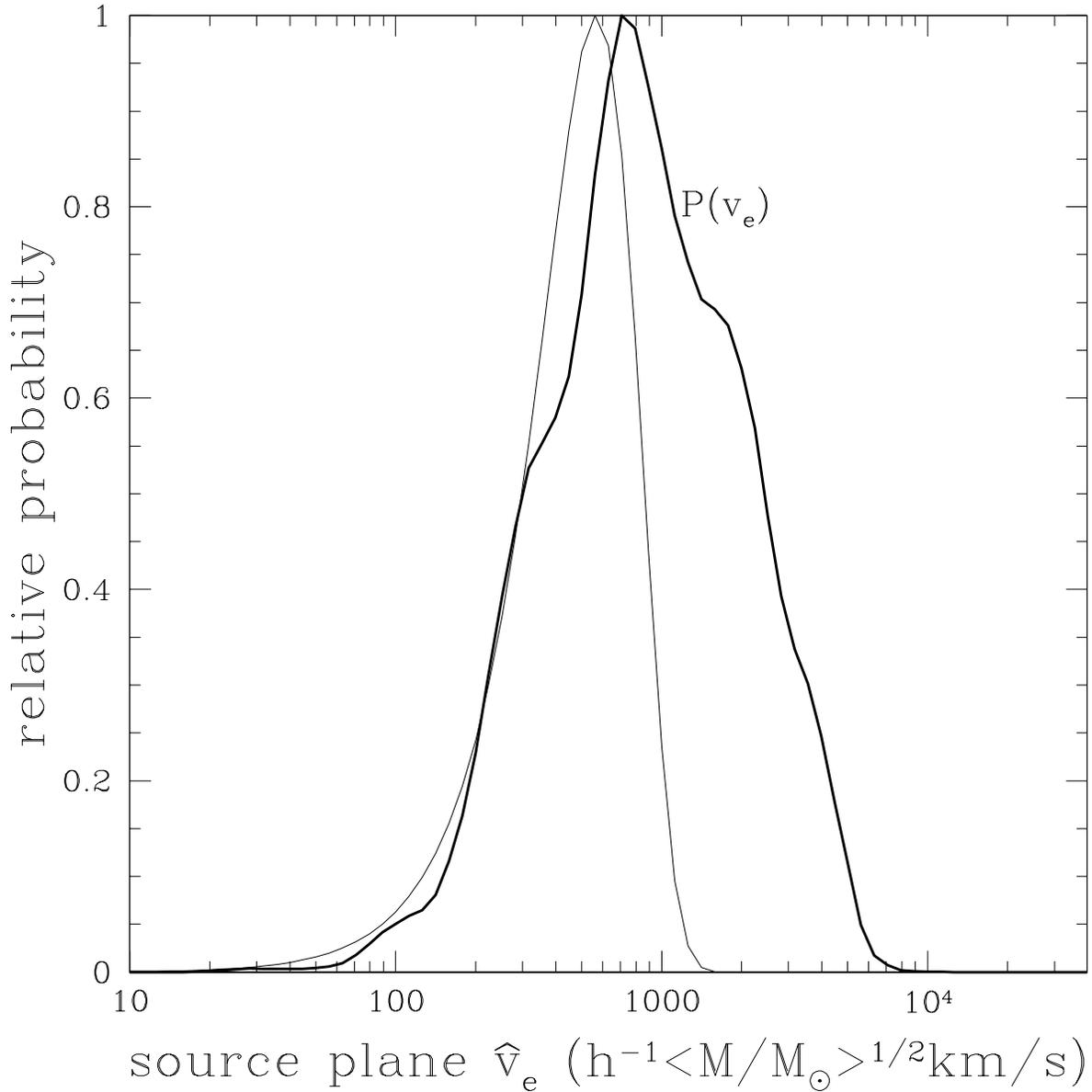}
\caption{Normalized probability distribution for the effective source plane velocity 
  ($\hat{v}_e$, heavy solid line) as compared to our estimated probability distribution
  for the true source plane effective velocity $v_e$ (light solid line).  Since the 
  average microlens mass $\protect{\avgm}$ is related to the two velocities 
  by $\hat{v}_e \propto v_{e}/\protect{\avgm}^{1/2}$,
  high (low) ratios of $\hat{v}_e/v_e$ correspond to low (high) mass microlenses. 
   \label{fig:Ve}}
\end{figure}

\begin{figure}
\epsscale{1.0}
\plotone{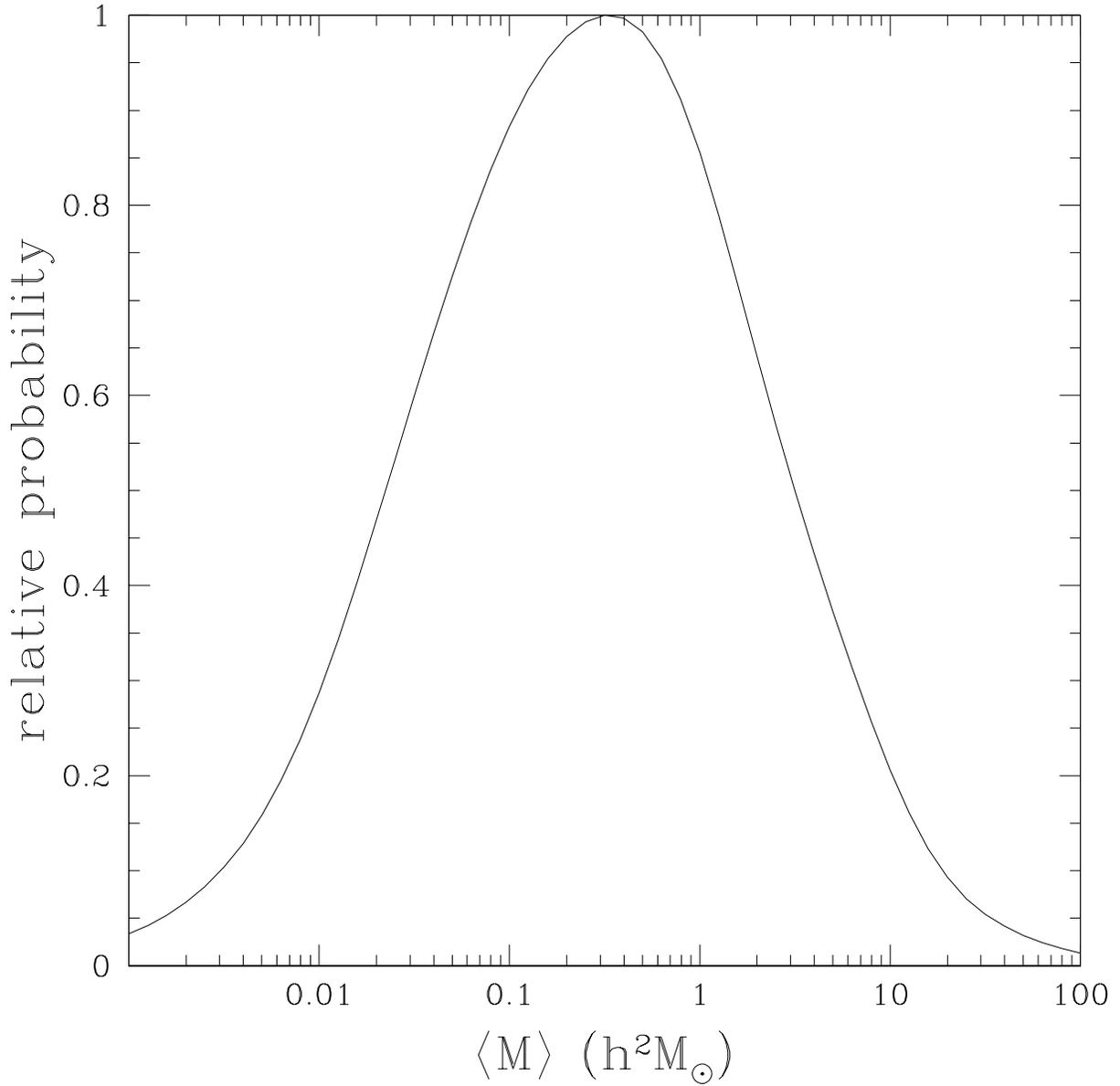}
\caption{Probability distribution for the average stellar mass $\protect{\avgm}$
	in the lens galaxy.  The uncertainty is relatively large because 
        $\protect{\avgm}\propto v_e^{-2}$, but it is consistent with normal
         stellar populations. \label{fig:mstar}}
\end{figure}

Figures~\ref{fig:Ve} and~\ref{fig:mstar} show the results for the system motions and
the lens galaxy stellar mass estimate.  The large flux ratio anomalies combined with the
limited amount of observed variability means that the effective velocity in the system
must be relatively low.  We find that 
$280 \, {\rm km \: s^{-1}} \lesssim \hat{v}_e \lesssim 749 \, {\rm km \: s^{-1}}$ (68\% confidence),
and this changes little if we adjust the D image light curve to be 1~mag fainter.
If we compare the effective velocity distribution to our model for the possible 
distribution of physical velocities (Fig.~\ref{fig:Ve}), we can estimate the 
mean microlens mass since the two velocities are related by 
$\hat{v}_e = v_e/\langle M/M_\odot\rangle^{1/2}$.  The broad range permitted
for $\hat{v}_e$ translates into a broad range for the stellar mass estimates, with
$0.02 \, {\rm M_\sun} \lesssim \avgm \lesssim 1.0 \, {\rm M_\sun}$ (68\% confidence).
The low mass solutions correspond to large sources with high effective velocities
and the high mass solutions correspond to small sources with low effective 
velocities.  

We were somewhat surprised to find that the present data do 
not distinguish between the lens models at all because we had based our 
expectations on the \citep{Schechter&Wambsganss2002} picture in which models with
low $\kappa_*/\kappa$ dominate the probability of finding a faint saddle point
image like D. We found instead that the probability distribution
for $f_{M/L}$ is basically flat. This result is
little affected by imposing the prior on the permitted mass range of the microlenses
or by adjusting the image D light curve to be 1~mag fainter. 
In comparison, the microlensing models for SDSS 0924+0219 by 
\citet{Keeton2006} strongly favored models with low  $\kappa_*/\kappa$. We
believe the differences between the results are 
due to our use of finite--sized sources, which 
significantly enhance the probability of large demagnifications relative to large
magnifications because the high magnification regions (caustics) are more affected 
by finite source sizes.  Another source of differences is that we are analyzing a 
more strongly constrained problem by requiring that the models fit the observed 
light curves rather than simply fit the instantaneous flux ratios.

\subsection{Quasar Structure}
\label{sec:quasarstructure}

\begin{figure}
\epsscale{1.0}
\plotone{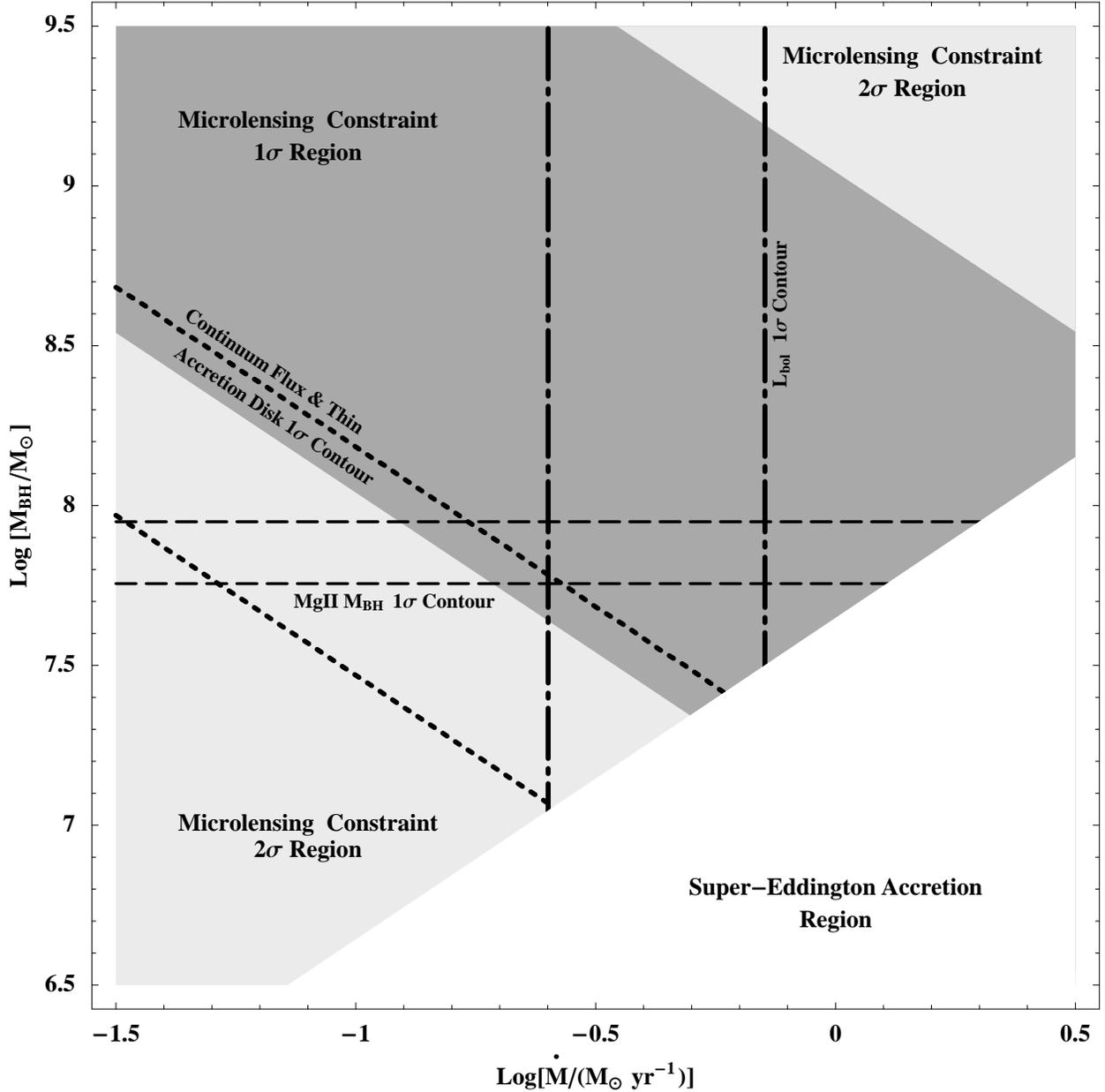}
\caption{Constraints on the mass and accretion rate of the SDSS0924+0219 quasar.
 The dashed horizontal lines show the estimated black hole mass based on the
 \ion{Mg}{2} 2800{\AA} emission line width.  The vertical dash-dotted lines 
 show the estimated accretion rate assuming a radiative efficiency of 
 10\% and the \citet{Kaspi2000} method for estimating the bolometric
 luminosity.  The dotted lines show the constraint on the product $M_{BH}\dot{M}$
 from the continuum flux and thin accretion disk theory.  The shaded
 regions show the constraint on the product $M_{BH}\dot{M}$ from the
 microlensing data with a prior of $0.1 \, M_\sun 
 < \langle M\rangle < 1.0 \, M_\odot$ on the mean mass of the microlenses.
 The accretion is super-Eddington in the lower right
 corner, so we terminated the constraints on the line where $L=L_{Edd}$.
All values plotted are scaled to $h=0.7$. 
\label{fig:mbh}}
\end{figure}

One objective of our monitoring program is to study the structure of quasar 
accretion disks.  We start by estimating the black hole mass using 
the ${\rm M_{BH}}$, \ion{Mg}{2} line width, luminosity relations of 
\citet{McLure2002} and \citet{Kollmeier2005}. We measured the \ion{Mg}{2}(2800{\AA}) 
line width in spectra obtained by \citet{Ofek2005} following the procedures 
detailed in \citet{Kollmeier2005}, and we estimated
the magnification--corrected continuum luminosity at 3000{\AA}, $L_{\lambda}$(3000{\AA}), 
using a power law fit to our HST data.  For the
\citet{McLure2002} calibration we find a black hole mass of
$M_{BH}=(6.3 \pm 1.5) \times 10^7 \: {\rm M_{\sun}}$, and for
the \citep{Kollmeier2005} calibration we find 
$M_{BH}=(7.3 \pm 2.4 )\times 10^7 \: {\rm M_{\sun}}$.  We adopt
the estimate based on the \citet{Kollmeier2005} calibration.  
Similarly, we estimate that the magnification--corrected bolometric 
luminosity of the quasar is   
$L_{bol}= (2.7\pm1.3) \times 10^{45} \: {\rm erg\;s^{-1}}$ 
where we follow  \citet{Kaspi2000} in assuming that
$L_{bol} \simeq 9 \times \lambda L_{\lambda}$(5100{\AA}). This bolometric
luminosity corresponds to an accretion rate $\dot{M}=(0.48\pm0.24)
\eta_{0.1}^{-1} {\rm M_{\sun}}$~yr$^{-1}$, where $\eta=0.1 \eta_{0.1}$
is the radiative efficiency of the accretion.   
Fig.~\ref{fig:mbh} summarizes these ``classical'' constraints on the
quasar.

The new constraint we obtain from the microlensing observations is on the
size $r_s$ of the quasar, which we can also estimate using our accretion
disk model and the observed flux.  A standard thin disk model 
\citep{Shakura1973} radiates as a black
body with a temperature profile of $T = T_s (R/r_s)^{-3/4}$, and the scale
length we measure should correspond to the point in the disk where the 
temperature corresponds to the rest-frame wavelength of the filter band
pass.  For our $R$-band data (2770{\AA} in the quasar rest frame), our scale
length corresponds to the point where $T_s(r_s)\simeq 5.2\times 10^4$~K.
If the viscous energy release is radiated locally and we are well removed from
the Schwarzschild radius, then the disk temperature and scale length are
related to the black hole mass $M_{BH}$ and accretion rate $\dot{M}$ by
\begin{equation}
    \sigma T_{s}^4 = 3 G M_{BH} \dot{M}/8\pi r_s^3
\end{equation} 
\citep{Shakura1973}, so a measurement of $r_s$ constrains the product
$M_{BH} \dot{M}$. One 
means of estimating $r_s$ is to simply compute what it must be to
produce the observed $R$-band flux.  Again assuming a standard,
face--on thin disk, the emission profile is  
\begin{equation}
   I(R) \propto \left[ \exp\left( (R/r_s)^{3/4} \right) -1 \right]^{-1}
\end{equation}
\citep{Shakura1973}.
Assuming that the disk radiates locally as a blackbody, we integrate this 
emission profile over the physical extent of the disk to estimate its specific 
luminosity $L_{\nu,em}$ in the rest frame.
Incorporating the system's geometry and correcting for redshift effects, 
we convert $L_{\nu,em}$ to $F_{\lambda,obs}$, the 
specific flux in the observed frame.
We then solve $F_{\lambda,obs}$ for $r_s$ to yield
\begin{equation}
r_{s_{\lambda,obs}} = 2.83 \times 10^{15} {1 \over \sqrt{\cos i}}
\left( { D_{OS} \over r_H } \right) \left( { \lambda_{obs} \over {\rm \micron} } \right)^{3/2}
10^{-0.2(M_{\lambda_{obs}}-19)} \: h^{-1} \: {\rm cm},
\label{eqn:fluxsize} 
\end{equation} 
where $M_{\lambda,obs}$ is the observed magnitude, 
$D_{OS}/r_H$ is the angular diameter distance to the quasar in units of the Hubble
radius and $i$ is the disk inclination angle, 
assumed to have an average value $\langle i \rangle = 60\degr$. 
For SDSS 0924+0219, we find an unmagnified {\it HST} $I$-band 
magnitude $I=21.24 \pm 0.25$~mag, yielding
the scale radius at the redshifted center of the $HST$ $I$-band, 
$r_{s_{I,obs}}=6.3 \pm 1.6 \times 10^{14} \: {\rm cm}$.
Assuming the $T^{-4/3}$ scaling of thin disk 
theory, we estimate an $R$-band disk size of 
$r_{s_{R,obs}} = 5.0 \pm 1.3 \times 10^{14} \: {\rm cm}$.
We show the resulting constraint on $M_{BH} \dot{M}$ in Fig.~\ref{fig:mbh}.

\begin{figure}
\epsscale{1.0}
\plotone{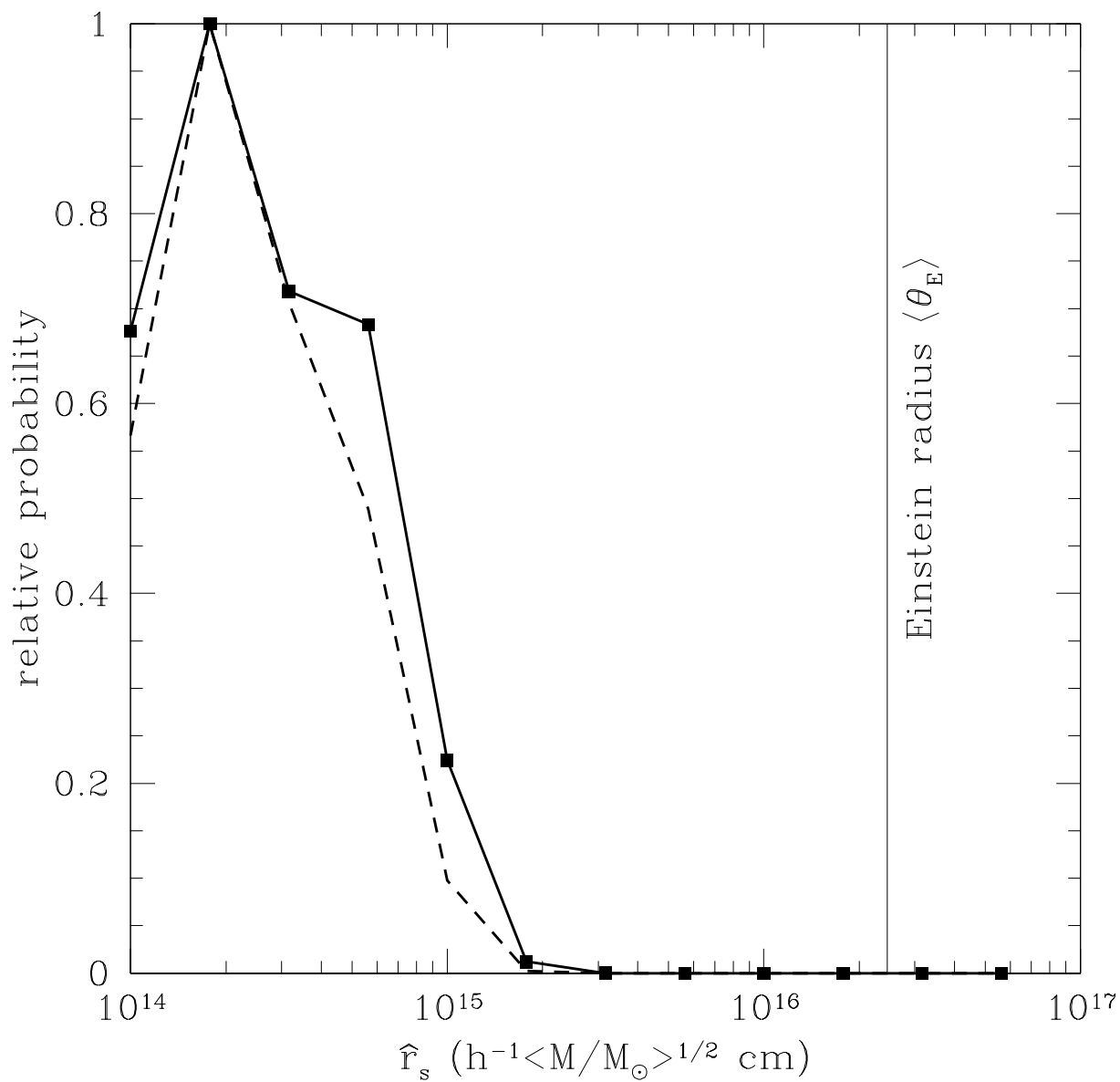}
\caption{Probability distributions for the scaled source size $\hat{r}_s$.  
  The heavy dashed line shows the estimate for $\hat{r}_s$ including 
  a prior of $0.1 \, M_\sun < \langle M\rangle < 1.0 \, M_\sun$ on
  the mass of the stars. The vertical line shows the Einstein Radius 
  $\langle\theta_E\rangle$ of the average mass star. 
  \label{fig:rs_einstein}}
\end{figure}

Microlensing tests this theoretical calculation because the amplitude of
the microlensing variability is controlled by the projected area of the 
source that smooths the magnification patterns.  Fig.~\ref{fig:rs_einstein}
shows our estimate of the scaled $R$-band source size $\hat{r}_s$, which
is related to the physical source size by 
$r_s = \hat{r}_s \langle M/M_\odot\rangle^{1/2}$.  The source must
be quite compact relative to the Einstein radius of the typical
microlens, with an exact bound that is presently difficult to 
determine because of the limited level of observed variability.
We face two technical problems in extending Fig.~\ref{fig:rs_einstein}
to smaller source sizes.  The first problem is that our analysis code
is presently limited to $8192^2$ magnification patterns, so when using
an outer dimension large enough to produce a reasonable statistical
representation of the magnification patterns it is difficult to resolve
such small scales.  The second problem is that even if we could resolve
the smallest scales, we would find that the probability distribution
flattens and becomes constant at small scale lengths.  This occurs 
because the differences between small smoothing lengths are detectable
only during caustic crossings -- if our light curves do not extend to
within a source size of a caustic, there is little effect from using a
still smaller source size.  Complete convergence at small scales will
require a light curve with caustic crossings. 

\begin{figure}
\epsscale{1.0}
\plotone{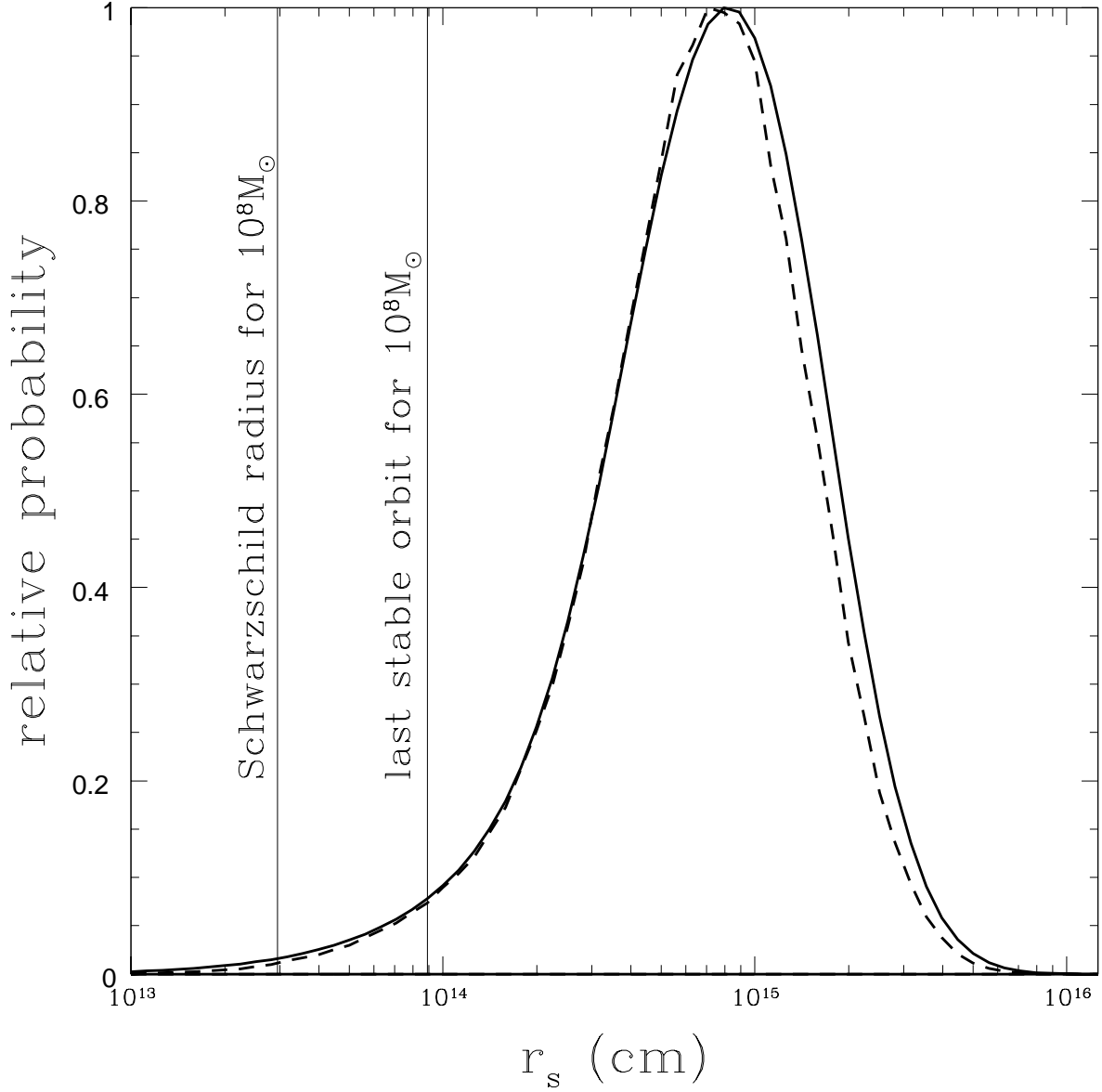}
\caption{Probability distributions for the physical source size $r_s$.
  The dashed curve shows the estimate for $r_s$ with a prior of 
  $0.1 \, M_\sun < \langle M\rangle < 1.0 \, M_\sun$ on
  the mean mass of the microlenses.  The vertical line shows the 
  Schwarzschild radius $R_{BH}=2GM_{BH}/c^2$ of a $10^8 M_\odot$ black hole.  
  The last stable orbit for a Schwarzschild black hole is
  at $3R_{BH}$.\label{fig:rs}}
\end{figure}

Despite these problems, we can estimate the physical source size of the
accretion disk reasonably well because there is a fairly strong degeneracy
between the scaled source size $\hat{r}_s$, the scaled velocity $\hat{v}_e$
and the microlens mass scale $\langle M/M_\odot\rangle$ in the sense that
more compact sources require smaller scaled velocities which implies a
larger microlens mass scale for the conversion to the physical source
size \citep[see][]{Kochanek2004}.  Fig.~\ref{fig:rs} shows the estimates of the
physical size $r_s$ both with and without the prior on the microlens
masses.  Reassuringly, the results depend only weakly on the prior. 
Nonetheless, will adopt the results with the mass 
prior: $3.0 \times 10^{14} \: {\rm cm} \lesssim r_s \lesssim 1.4 
\times 10^{15} \: {\rm cm}$ at 68\% confidence.  This is consistent
with our earlier estimate from the continuum flux but is a weaker constraint
on $M_{BH} \dot{M}$, as shown in Fig.~\ref{fig:mbh}.

\subsection{Expectations for the Future Behavior of SDSS0924+0219}
\label{sec:expectations}

\begin{figure}
\epsscale{1.0}
\plotone{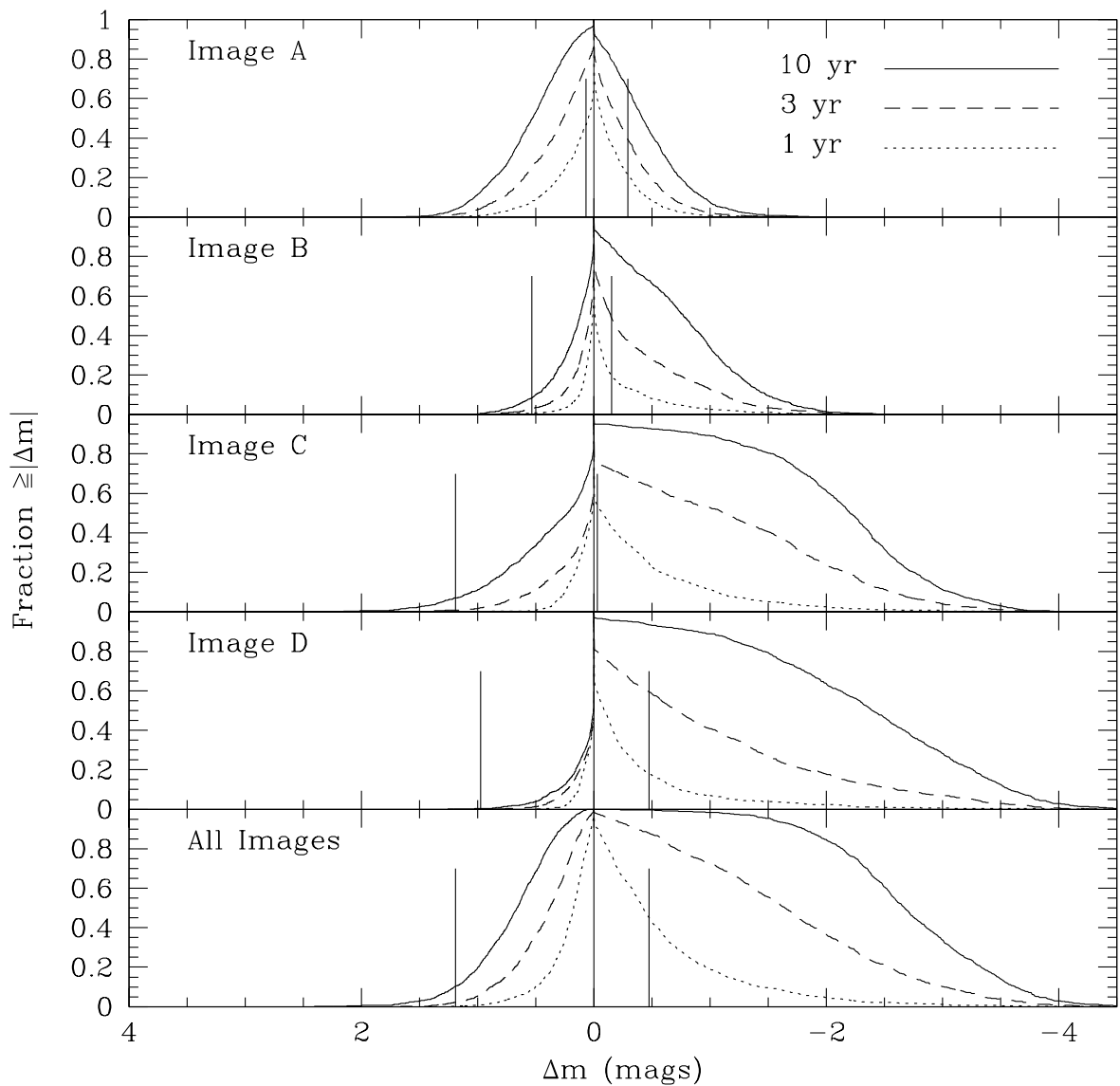}
\caption{Fraction of light curves 
with $\chi^2/N_{DOF}<2.3$ that undergo a change in magnification larger than 
$\Delta m$ magnitudes towards either brighter (right) or fainter (left) fluxes after one
(dotted), three (dashed) or ten (solid) years. The vertical lines show the largest
observed $\Delta m$ for our present light curves.
\label{fig:deltaplots}}
\end{figure}

For each light curve which passed a threshold of $\chi^2/N_{DOF} \leq 2.3$, we generated a 
light curve for a period of ten years beyond our most recent observation. For each image,
we then tracked the maximum change in the brightness in both the positive and negative
directions on 1, 3 and 10 year time scales.  The normalized, cumulative distributions
of these maximum changes are shown in Fig.~\ref{fig:deltaplots}. One of the more obvious 
predictions of Fig.~\ref{fig:deltaplots} is that images A and B are likely to remain constant 
while images C and D are likely to become brighter. One of the original 
motivations of this study was to estimate the time scale on which
the flux ratio anomaly would vanish as D moved out of a low--magnification region
and became brighter.  Here we find an approximately 12\% chance that it will brighten
by at least a factor of 2 in the course of the next year and a roughly 45\% chance that
it will brighten by more than an order of magnitude over the next decade. For 
the separate calculation in which we lowered the flux of all points on the 
image D light curve by +1 magnitude, we predict a 9\% probability 
of image D brightening by a factor of two during the next year, 
and a 53\% chance that image D will brighten by a factor of ten during the next ten years. 

Our expectation that D brightening would be the means of solving the anomaly
was based on the preconception that D was a de-magnified saddle point in a
model with a small ratio between the stellar and total surface densities
$\kappa_*/\kappa$.  \citet{Schechter&Wambsganss2002} demonstrated that in
this regime there is an appreciable probability of strongly de-magnifying
saddle point images like D. Our results confirm this finding.
 
\section{Conclusions}
\label{sec:conclusions}

During the course of our monitoring campaign we have observed uncorrelated
variability in the four images of SDSS 0924+0219, evidence that microlensing
is occurring in this system. Furthermore, our models demonstrate
that microlensing is a viable explanation for the system's anomalous flux ratios.
This study does not rule out the alternative hypothesis that dark matter substructure
contributes to the anomaly, but it does firmly establish the presence of
microlensing variability and the ability of microlensing to explain the anomaly.  
As we continue to monitor SDSS 0924+0219, we expect to eventually measure the 
time delay, thereby restricting the range of permissible halo models, and to 
steadily reduce the uncertainties in the estimated mean stellar mass, accretion 
disk structure and black hole mass.  At some point over the next few years, we
should also see dramatic changes in the fluxes of the merging images.

We can also begin to compare microlensing estimates of the structure of quasar 
accretion disks.  In our original study \citep{Kochanek2004}, we modeled the 
significantly more luminous, but very similar redshift, quasar Q2237+0305 
($M_V = -25.8 \pm 0.5$ versus $M_V = -21.7 \pm 0.7$ 
after correcting for magnification).  As we would expect from 
accretion disk theory, the microlensing analyses indicate that the more 
luminous quasar has a significantly larger scale ($r_s \simeq 
4.1 \times 10^{15} \: {\rm cm}$ 
versus $r_s \simeq 6.9 \times 10^{14} \: {\rm cm}$) 
and black hole mass ($M_{BH} \simeq
1.1 \times 10^9 \, {\rm M_{\sun}}$  versus $1.3 \times 10^8 \, {\rm M_{\sun}}$).  The next 
step is to combine the microlensing analyses of many lenses to explore these 
correlations in detail and to use the wavelength dependence of the microlensing
variability to study the structure of individual disks.  This 
next step should be possible very shortly.

\acknowledgements We thank E. Turner for providing data from APO and P.
Schechter and W. Barkhouse for the data from Magellan. We thank E. Agol
for discovering an error in the original manuscript.
We thank C. Onken, B. Peterson and R. Pogge for discussions about
quasar structure and black hole mass estimation from emission line widths.
This research made extensive use
of a Beowulf computer cluster obtained through the Cluster Ohio
program of the Ohio Supercomputer Center. Support for program HST-GO-9744 was
provided by NASA through a grant from the Space Telescope Science Institute, which 
is operated by the Association of Universities for Research in Astronomy, Inc., under
NASA contract NAS-5-26666.

{\it Facilities:} \facility{HST (NICMOS, ACS)}, 
\facility{Hiltner (EIGHTK, Echelle, RETROCAM)}, \facility{ARC (Spicam)}, 
\facility{WIYN (WTTM)}, \facility{CTIO:2MASS (ANDICAM)}
	

\clearpage

\begin{deluxetable}{cccccc}
\tablecaption{HST Astrometry and Photometry of SDSS0924+0219 }
\tablehead{Component &\multicolumn{2}{c}{Astrometry}
                 &\multicolumn{3}{c}{Photometry}\\
                 \colhead{}
                 &\colhead{$\Delta\hbox{RA}$}
                 &\colhead{$\Delta\hbox{Dec}$}
                 &\colhead{H=F160W}
                 &\colhead{I=F814W}
                 &\colhead{V=F555W}
                 }

\startdata
A  &$\equiv 0$ &$\equiv 0$ &$17.96\pm0.02$ &$18.77\pm0.05$ &$19.61\pm0.01$\\
B  &$+0\farcs061\pm0\farcs003$ &$-1\farcs805\pm0\farcs003$ &$18.87\pm0.03$ &$19.64\pm0.07$ &$20.36\pm0.05$\\
C  &$-0\farcs968\pm0\farcs003$ &$-0\farcs676\pm0\farcs005$ &$19.22\pm0.02$ &$20.22\pm0.09$ &$20.74\pm0.09$\\
D  &$+0\farcs536\pm0\farcs003$ &$-0\farcs427\pm0\farcs003$ &$20.64\pm0.06$ &$22.00\pm0.22$ &$22.94\pm0.13$\\
G  &$-0\farcs183\pm0\farcs004$ &$-0\farcs858\pm0\farcs004$ &$17.23\pm0.04$ &$19.39\pm0.06$ &$20.83\pm0.03$\\
\enddata
\label{tab:hstastrom_phot}
\end{deluxetable}

\begin{deluxetable}{ccccccccccccc}
\tabletypesize{\scriptsize}
\tablewidth{0pt}
\tablecaption{Macroscopic Lens Mass Models}

\tablehead{$f_{M/L}$
                &\multicolumn{4}{c}{Convergence $\kappa$}
                &\multicolumn{4}{c}{Shear $\gamma$}
                &\multicolumn{4}{c}{$\kappa_{*}/\kappa$}\\
		\colhead{}
		&\colhead{A}
                &\colhead{B}
                &\colhead{C}
                &\colhead{D}
                &\colhead{A}
                &\colhead{B}
		&\colhead{C}
		&\colhead{D}
		&\colhead{A}
                &\colhead{B}
                &\colhead{C}
                &\colhead{D}	
                }

\startdata
$0.1$ &$ 0.74 $&$ 0.71 $&$ 0.77 $&$ 0.75 $&$ 0.24 $&$ 0.21 $&$ 0.29 $&$ 0.28 $&$ 0.020 $&$ 0.018 $&$ 0.024 $&$ 0.021$ \\
$0.2$ &$ 0.67 $&$ 0.65 $&$ 0.70 $&$ 0.67 $&$ 0.30 $&$ 0.26 $&$ 0.37 $&$ 0.36 $&$ 0.042 $&$ 0.038 $&$ 0.050 $&$ 0.044$ \\
$0.3$ &$ 0.61 $&$ 0.59 $&$ 0.64 $&$ 0.61 $&$ 0.35 $&$ 0.30 $&$ 0.44 $&$ 0.43 $&$ 0.061 $&$ 0.055 $&$ 0.072 $&$ 0.062$ \\
$0.4$ &$ 0.53 $&$ 0.51 $&$ 0.56 $&$ 0.53 $&$ 0.42 $&$ 0.36 $&$ 0.53 $&$ 0.52 $&$ 0.11 $&$ 0.10 $&$ 0.13 $&$ 0.11$ \\
$0.5$ &$ 0.47 $&$ 0.45 $&$ 0.49 $&$ 0.46 $&$ 0.48 $&$ 0.41 $&$ 0.61 $&$ 0.60 $&$ 0.15 $&$ 0.14 $&$ 0.17 $&$ 0.15$ \\
$0.6$ &$ 0.40 $&$ 0.38 $&$ 0.44 $&$ 0.41 $&$ 0.54 $&$ 0.45 $&$ 0.69 $&$ 0.67 $&$ 0.19 $&$ 0.18 $&$ 0.23 $&$ 0.20$ \\
$0.7$ &$ 0.31 $&$ 0.29 $&$ 0.35 $&$ 0.31 $&$ 0.62 $&$ 0.51 $&$ 0.80 $&$ 0.77 $&$ 0.34 $&$ 0.32 $&$ 0.38 $&$ 0.34$ \\
$0.8$ &$ 0.25 $&$ 0.23 $&$ 0.29 $&$ 0.26 $&$ 0.67 $&$ 0.55 $&$ 0.88 $&$ 0.84 $&$ 0.45 $&$ 0.42 $&$ 0.50 $&$ 0.46$ \\
$0.9$ &$ 0.17 $&$ 0.16 $&$ 0.22 $&$ 0.18 $&$ 0.73 $&$ 0.61 $&$ 0.96 $&$ 0.92 $&$ 0.72 $&$ 0.70 $&$ 0.77 $&$ 0.73$ \\
$1.0$ &$ 0.13 $&$ 0.12 $&$ 0.17 $&$ 0.13 $&$ 0.77 $&$ 0.64 $&$ 1.02 $&$ 0.98 $&$ 1.00 $&$ 1.00 $&$ 1.00 $&$ 1.00$ \\

\enddata
\tablecomments{Convergence $\kappa$, shear $\gamma$ and the fraction of the total surface density 
composed of stars $\kappa_{*}/\kappa$ at each image location for the series of 
macroscopic mass models.}
\label{tab:macromodels}
\end{deluxetable}

\def\hm{\hphantom{-}}
\begin{deluxetable}{ccccccccc}
\tabletypesize{\scriptsize}
\tablecaption{SDSS0924+0219 Light curves}
\tablewidth{0pt}
\tablehead{ HJD &\multicolumn{1}{c}{$\chi^2/N_{dof}$}
		&\multicolumn{1}{c}{Seeing}
                &\multicolumn{1}{c}{QSO A} &\multicolumn{1}{c}{QSO B} 
                &\multicolumn{1}{c}{QSO C} &\multicolumn{1}{c}{QSO D}
                &\multicolumn{1}{c}{$\langle\hbox{Stars}\rangle$}
                &\multicolumn{1}{c}{Source} 
              }
\startdata
$2957.814$ &$  0.70$ &$1\farcs22$ &$ 2.812\pm 0.028$ &$ 3.515\pm 0.033$ &$ 3.883\pm 0.054$ &$ 5.041\pm 0.217$ &$-0.032\pm 0.003$ &SMARTS \\ 
$2964.789$ &$  0.65$ &$1\farcs42$ &$ 2.864\pm 0.024$ &$ 3.516\pm 0.024$ &$ 3.895\pm 0.042$ &$ 4.840\pm 0.144$ &$\hm 0.004\pm 0.003$ &SMARTS \\ 
$2976.775$ &$  0.64$ &$1\farcs31$ &$ 2.800\pm 0.020$ &$ 3.464\pm 0.022$ &$ 3.892\pm 0.039$ &$ 5.006\pm 0.149$ &$-0.001\pm 0.003$ &SMARTS \\ 
$2984.721$ &$  0.56$ &$1\farcs20$ &$ 2.811\pm 0.036$ &$ 3.453\pm 0.040$ &$ 3.852\pm 0.068$ &$ 4.632\pm 0.189$ &$-0.037\pm 0.003$ &SMARTS \\ 
$2991.791$ &$  0.79$ &$1\farcs08$ &$ 2.766\pm 0.016$ &$ 3.531\pm 0.019$ &$ 3.991\pm 0.033$ &$ 5.138\pm 0.128$ &$\hm 0.012\pm 0.003$ &SMARTS \\ 
$2995.932$ &$  1.82$ &$1\farcs20$ &$ 2.785\pm 0.012$ &$ 3.507\pm 0.013$ &$ 3.928\pm 0.022$ &$ 5.147\pm 0.105$ &$\hm 0.084\pm 0.002$ &MDM--EIGHTK \\ 
$2998.787$ &$  1.31$ &$0\farcs94$ &$ 2.804\pm 0.014$ &$ 3.541\pm 0.017$ &$ 4.047\pm 0.029$ &$ 5.258\pm 0.120$ &$\hm 0.027\pm 0.003$ &SMARTS \\ 
$3022.797$ &$  1.82$ &$1\farcs09$ &$ 2.882\pm 0.019$ &$ 3.417\pm 0.018$ &$ 4.161\pm 0.037$ &$ 4.937\pm 0.116$ &$\hm 0.015\pm 0.003$ &SMARTS \\ 
$3029.751$ &$  2.13$ &$0\farcs90$ &$ 2.865\pm 0.014$ &$ 3.386\pm 0.015$ &$ 4.187\pm 0.029$ &$ 5.092\pm 0.092$ &$\hm 0.021\pm 0.003$ &SMARTS \\ 
$3036.673$ &$  0.70$ &$0\farcs99$ &$ 2.780\pm 0.016$ &$ 3.441\pm 0.019$ &$ 4.216\pm 0.042$ &$ 5.209\pm 0.144$ &$\hm 0.004\pm 0.003$ &SMARTS \\ 
$3047.794$ &$  0.94$ &$1\farcs31$ &$ 2.749\pm 0.024$ &$ 3.408\pm 0.024$ &$ 4.331\pm 0.063$ &$ 5.033\pm 0.193$ &$-0.013\pm 0.003$ &SMARTS \\ 
$3055.689$ &$  0.68$ &$1\farcs13$ &$ 2.819\pm 0.009$ &$ 3.555\pm 0.012$ &$ 4.206\pm 0.020$ &$ 5.353\pm 0.071$ &$-0.025\pm 0.002$ &MAGELLAN \\ 
$3056.744$ &$  0.80$ &$1\farcs11$ &$ 2.757\pm 0.016$ &$ 3.428\pm 0.017$ &$ 4.201\pm 0.036$ &$ 5.152\pm 0.133$ &$\hm 0.016\pm 0.003$ &SMARTS \\ 
$3064.680$ &$  1.30$ &$1\farcs08$ &$ 2.836\pm 0.018$ &$ 3.381\pm 0.018$ &$ 4.187\pm 0.039$ &$ 4.948\pm 0.119$ &$\hm 0.010\pm 0.003$ &SMARTS \\ 
$3065.805$ &$  0.61$ &$1\farcs28$ &$ 2.832\pm 0.018$ &$ 3.389\pm 0.017$ &$ 4.179\pm 0.039$ &$ 4.560\pm 0.091$ &$\hm 0.080\pm 0.002$ &MDM--ECHELLE \\ 
$3071.649$ &$  0.53$ &$1\farcs08$ &$ 2.800\pm 0.033$ &$ 3.423\pm 0.039$ &$ 4.269\pm 0.092$ &$ 4.782\pm 0.206$ &$-0.044\pm 0.003$ &SMARTS \\ 
$3075.774$ &$  4.24$ &$1\farcs07$ &$ 2.828\pm 0.012$ &$ 3.491\pm 0.013$ &$ 4.005\pm 0.022$ &$ 5.046\pm 0.091$ &$\hm 0.053\pm 0.002$ &MDM--EIGHTK \\ 
$3078.586$ &$  0.57$ &$1\farcs14$ &$ 2.791\pm 0.019$ &$ 3.446\pm 0.020$ &$ 4.234\pm 0.045$ &$ 4.888\pm 0.128$ &$\hm 0.007\pm 0.003$ &SMARTS \\ 
$3080.834$ &$  0.56$ &$1\farcs31$ &$ 2.797\pm 0.017$ &$ 3.419\pm 0.016$ &$ 4.268\pm 0.041$ &$ 4.742\pm 0.098$ &$-0.002\pm 0.003$ &  APO \\ 
$3087.648$ &$  1.49$ &$0\farcs86$ &$ 2.770\pm 0.007$ &$ 3.408\pm 0.008$ &$ 4.294\pm 0.012$ &$ 5.384\pm 0.043$ &$\hm 0.188\pm 0.003$ &WIYN--WTTM \\ 
$3089.599$ &$  1.11$ &$1\farcs04$ &$ 2.781\pm 0.016$ &$ 3.409\pm 0.017$ &$ 4.198\pm 0.036$ &$ 5.150\pm 0.126$ &$\hm 0.011\pm 0.003$ &SMARTS \\ 
$3090.666$ &$  0.79$ &$1\farcs50$ &$ 2.833\pm 0.018$ &$ 3.336\pm 0.013$ &$ 4.450\pm 0.041$ &$ 4.372\pm 0.066$ &$\hm 0.168\pm 0.003$ &WIYN--WTTM \\ 
$3101.623$ &$  0.60$ &$1\farcs20$ &$ 2.732\pm 0.033$ &$ 3.417\pm 0.037$ &$ 4.420\pm 0.103$ &$ 4.567\pm 0.179$ &$-0.049\pm 0.003$ &SMARTS \\ 
$3108.511$ &$  2.18$ &$1\farcs02$ &$ 2.710\pm 0.016$ &$ 3.361\pm 0.018$ &$ 4.213\pm 0.039$ &$ 5.063\pm 0.129$ &$\hm 0.009\pm 0.003$ &SMARTS \\ 
$3115.562$ &$  1.11$ &$1\farcs19$ &$ 2.697\pm 0.019$ &$ 3.398\pm 0.020$ &$ 4.273\pm 0.049$ &$ 4.889\pm 0.140$ &$\hm 0.003\pm 0.003$ &SMARTS \\ 
$3127.473$ &$  0.88$ &$0\farcs92$ &$ 2.702\pm 0.020$ &$ 3.364\pm 0.025$ &$ 4.249\pm 0.058$ &$ 4.735\pm 0.133$ &$-0.022\pm 0.003$ &SMARTS \\ 
$3134.453$ &$  0.79$ &$0\farcs91$ &$ 2.631\pm 0.021$ &$ 3.467\pm 0.031$ &$ 4.442\pm 0.080$ &$ 4.861\pm 0.167$ &$-0.035\pm 0.003$ &SMARTS \\ 
$3141.510$ &$  0.82$ &$1\farcs10$ &$ 2.653\pm 0.021$ &$ 3.467\pm 0.025$ &$ 4.348\pm 0.060$ &$ 4.638\pm 0.131$ &$-0.018\pm 0.003$ &SMARTS \\ 
$3159.468$ &$  0.69$ &$1\farcs07$ &$ 2.596\pm 0.024$ &$ 3.484\pm 0.035$ &$ 4.382\pm 0.087$ &$ 5.053\pm 0.238$ &$-0.044\pm 0.003$ &SMARTS \\ 
$3169.462$ &$  0.79$ &$1\farcs28$ &$ 2.635\pm 0.020$ &$ 3.582\pm 0.025$ &$ 4.269\pm 0.057$ &$ 5.135\pm 0.200$ &$-0.012\pm 0.003$ &SMARTS \\ 
$3324.813$ &$  0.42$ &$1\farcs44$ &$ 2.675\pm 0.030$ &$ 3.882\pm 0.047$ &$ 4.450\pm 0.099$ &$ 5.528\pm 0.404$ &$-0.032\pm 0.003$ &SMARTS \\ 
$3327.800$ &$  0.74$ &$1\farcs46$ &$ 2.740\pm 0.027$ &$ 3.960\pm 0.042$ &$ 4.429\pm 0.083$ &$ 5.217\pm 0.265$ &$-0.016\pm 0.003$ &SMARTS \\ 
$3330.766$ &$  0.74$ &$1\farcs45$ &$ 2.776\pm 0.037$ &$ 4.050\pm 0.067$ &$ 4.466\pm 0.118$ &$ 5.129\pm 0.317$ &$-0.038\pm 0.003$ &SMARTS \\ 
$3344.723$ &$  0.89$ &$1\farcs43$ &$ 2.767\pm 0.026$ &$ 3.923\pm 0.039$ &$ 4.628\pm 0.096$ &$ 5.754\pm 0.399$ &$-0.018\pm 0.003$ &SMARTS \\ 
$3349.758$ &$  0.56$ &$1\farcs49$ &$ 2.703\pm 0.032$ &$ 3.953\pm 0.048$ &$ 4.971\pm 0.157$ &$ 5.441\pm 0.384$ &$-0.020\pm 0.003$ &SMARTS \\ 
$3357.714$ &$  1.08$ &$1\farcs26$ &$ 2.765\pm 0.021$ &$ 3.869\pm 0.029$ &$ 4.586\pm 0.069$ &$ 5.674\pm 0.297$ &$\hm 0.003\pm 0.003$ &SMARTS \\ 
$3360.767$ &$  1.57$ &$1\farcs14$ &$ 2.799\pm 0.020$ &$ 3.841\pm 0.026$ &$ 4.544\pm 0.056$ &$ 5.306\pm 0.192$ &$\hm 0.007\pm 0.003$ &SMARTS \\ 
$3376.734$ &$  1.05$ &$1\farcs15$ &$ 2.780\pm 0.019$ &$ 3.926\pm 0.029$ &$ 4.533\pm 0.058$ &$ 5.379\pm 0.202$ &$\hm 0.006\pm 0.003$ &SMARTS \\ 
$3382.701$ &$  0.61$ &$1\farcs17$ &$ 2.772\pm 0.022$ &$ 4.012\pm 0.039$ &$ 4.674\pm 0.082$ &$ 5.468\pm 0.255$ &$-0.017\pm 0.003$ &SMARTS \\ 
$3394.741$ &$  0.61$ &$0\farcs98$ &$ 2.779\pm 0.030$ &$ 3.919\pm 0.064$ &$ 4.908\pm 0.168$ &$ 5.616\pm 0.426$ &$-0.044\pm 0.003$ &SMARTS \\ 
$3403.751$ &$  1.00$ &$0\farcs98$ &$ 2.801\pm 0.018$ &$ 3.934\pm 0.028$ &$ 4.733\pm 0.062$ &$ 5.107\pm 0.141$ &$\hm 0.003\pm 0.003$ &SMARTS \\ 
$3410.781$ &$  0.99$ &$1\farcs13$ &$ 2.767\pm 0.019$ &$ 3.905\pm 0.028$ &$ 4.681\pm 0.066$ &$ 5.257\pm 0.183$ &$-0.001\pm 0.003$ &SMARTS \\ 
$3427.737$ &$  0.55$ &$1\farcs32$ &$ 2.705\pm 0.041$ &$ 4.025\pm 0.085$ &$ 5.074\pm 0.247$ &$ 5.181\pm 0.399$ &$-0.031\pm 0.003$ &SMARTS \\ 
$3435.698$ &$  1.29$ &$0\farcs99$ &$ 2.793\pm 0.019$ &$ 3.872\pm 0.029$ &$ 4.666\pm 0.063$ &$ 5.155\pm 0.158$ &$-0.013\pm 0.003$ &SMARTS \\ 
$3443.698$ &$  1.45$ &$1\farcs29$ &$ 2.783\pm 0.024$ &$ 3.900\pm 0.033$ &$ 4.692\pm 0.080$ &$ 4.858\pm 0.161$ &$-0.006\pm 0.003$ &SMARTS \\ 
\cline{1-9}\\
$3450.662$ &$  0.66$ &$1\farcs20$ &$ 2.701\pm 0.038$ &$ 3.763\pm 0.066$ &$ 5.006\pm 0.224$ &$ 6.019\pm 0.807$ &$-0.031\pm 0.003$ &SMARTS \\ 
$3461.643$ &$  0.62$ &$1\farcs46$ &$ 2.703\pm 0.029$ &$ 3.953\pm 0.045$ &$ 4.965\pm 0.141$ &$ 5.626\pm 0.424$ &$-0.025\pm 0.003$ &SMARTS \\ 
$3468.550$ &$  0.95$ &$1\farcs28$ &$ 2.766\pm 0.024$ &$ 3.930\pm 0.035$ &$ 4.732\pm 0.087$ &$ 5.106\pm 0.200$ &$-0.020\pm 0.003$ &SMARTS \\ 
$3477.514$ &$  0.49$ &$1\farcs08$ &$ 2.676\pm 0.024$ &$ 3.846\pm 0.044$ &$ 4.604\pm 0.099$ &$ 5.900\pm 0.482$ &$-0.033\pm 0.003$ &SMARTS \\ 
$3490.481$ &$  0.77$ &$1\farcs06$ &$ 2.670\pm 0.019$ &$ 3.825\pm 0.030$ &$ 4.689\pm 0.073$ &$ 4.799\pm 0.136$ &$-0.010\pm 0.003$ &SMARTS \\ 
$3502.497$ &$  1.13$ &$1\farcs06$ &$ 2.632\pm 0.020$ &$ 3.776\pm 0.031$ &$ 4.662\pm 0.075$ &$ 4.938\pm 0.161$ &$-0.013\pm 0.003$ &SMARTS \\ 
$3520.515$ &$  0.85$ &$1\farcs34$ &$ 2.520\pm 0.025$ &$ 3.763\pm 0.038$ &$ 4.739\pm 0.112$ &$ 5.261\pm 0.306$ &$-0.024\pm 0.003$ &SMARTS \\ 
$3528.450$ &$  0.77$ &$1\farcs13$ &$ 2.587\pm 0.027$ &$ 3.743\pm 0.048$ &$ 4.901\pm 0.152$ &$ 5.322\pm 0.343$ &$-0.034\pm 0.003$ &SMARTS \\ 
$3676.836$ &$  0.50$ &$1\farcs40$ &$ 2.185\pm 0.020$ &$ 3.529\pm 0.033$ &$ 4.584\pm 0.114$ &$ 5.174\pm 0.306$ &$-0.030\pm 0.003$ &SMARTS \\ 
$3686.836$ &$  0.77$ &$1\farcs22$ &$ 2.120\pm 0.015$ &$ 3.518\pm 0.025$ &$ 4.375\pm 0.066$ &$ 5.295\pm 0.258$ &$-0.014\pm 0.003$ &SMARTS \\ 
$3696.825$ &$  0.54$ &$1\farcs31$ &$ 2.101\pm 0.025$ &$ 3.614\pm 0.059$ &$ 4.354\pm 0.140$ &$ 5.088\pm 0.403$ &$-0.043\pm 0.003$ &SMARTS \\ 
$3704.817$ &$  0.76$ &$1\farcs13$ &$ 2.184\pm 0.014$ &$ 3.520\pm 0.022$ &$ 4.364\pm 0.057$ &$ 4.969\pm 0.164$ &$-0.006\pm 0.003$ &SMARTS \\ 
$3711.988$ &$  1.35$ &$0\farcs99$ &$ 2.221\pm 0.007$ &$ 3.538\pm 0.008$ &$ 3.909\pm 0.011$ &$ 5.206\pm 0.046$ &$\hm 0.180\pm 0.003$ &MDM--RETROCAM \\ 
$3712.791$ &$  1.23$ &$1\farcs03$ &$ 2.213\pm 0.013$ &$ 3.505\pm 0.020$ &$ 4.220\pm 0.044$ &$ 5.099\pm 0.162$ &$-0.001\pm 0.003$ &SMARTS \\ 
$3719.804$ &$  0.54$ &$1\farcs13$ &$ 2.173\pm 0.019$ &$ 3.586\pm 0.041$ &$ 4.506\pm 0.109$ &$ 5.092\pm 0.279$ &$-0.041\pm 0.003$ &SMARTS \\ 

\enddata
\tablecomments{HJD is the Heliocentric Julian Day -- 2450000 days.
The goodness of fit of the image, $\chi^2/N_{dof}$, is used to rescale the
formal uncertainties when greater than unity (see text).   The QSO A-D
  columns give the magnitudes of the quasar images relative to the
  comparison stars. The $\langle\hbox{Stars}\rangle$ column gives the
  mean magnitude of the standard stars for that epoch relative to
  their mean for all epochs. }
\label{tab:lightcurves}
\end{deluxetable}

\clearpage

\end{document}